\documentstyle[12pt,epsf]{article}
\newlength{\dinwidth}
\newlength{\dinmargin}
\newcommand{\ba}{\begin{array}}
\newcommand{\ea}{\end{array}}
\newcommand{\be}{\begin{eqnarray}}
\newcommand{\ee}{\end{eqnarray}}

\newcommand{\tr}{\mbox{Tr}}

\newcommand{\F}{\bf F}
\newcommand{\G}{\bf G}

\begin{document}
\thispagestyle{empty}
\addtocounter{page}{-1}
\begin{flushright}
SNUST-000602\\
{\tt hep-th/0007089}
\end{flushright}
\vspace*{1.3cm}
\centerline{\Large \bf S-Duality, Noncritical Open String}
\vspace*{0.3cm}
\centerline{\Large \bf and}
\vspace*{0.3cm}
\centerline{\Large \bf Noncommutative Gauge Theory
~\footnote{Work supported in part by BK-21 Initiative in Physics (SNU
Team-2), KRF International Collaboration Grant, KOSEF 
Interdisciplinary Research Grant 98-07-02-07-01-5, and KOSEF Leading
Scientist Program 2000-1-11200-001-1. The work of RvU was
supported by the Czech Ministry of Education under contract
No. 144310006 and by the Swedish Institute.}}
\vspace*{1.2cm} 
\centerline{\bf Soo-Jong Rey${}^a$ {\rm and} Rikard von Unge${}^b$}
\vspace*{0.8cm}
\centerline{\it School of Physics \& Center for Theoretical Physics}
\vspace*{0.28cm}
\centerline{\it Seoul National University, Seoul 151-742 Korea ${}^a$}
\vspace*{0.4cm}
\centerline{\it  Department of Theoretical Physics \& Astrophysics }
\vspace*{0.28cm}
\centerline{\it Masaryk University, Brno CZ-61197 Czech Republic ${}^b$}
\vspace*{1.9cm}
\centerline{\bf abstract}
\vspace*{0.5cm}
We examine several aspects of S-duality of four-dimensional noncommutative 
gauge theory. By making duality transformation explicit, we find that S-dual 
of noncommutative gauge theory is defined in terms of dual noncommutative 
deformation. In `magnetic' noncommutative U(1) gauge theory, we show that,
in addition to gauge bosons, open D-strings constitute important low-energy 
excitations: noncritical open D-strings. Upon S-duality, they are mapped to 
noncritical open F-strings. We propose that, in dual `electric' noncommutative 
U(1) gauge theory, the latters are identified with gauge-invariant, open Wilson 
lines. We argue that the open Wilson lines are chiral due to underlying parity
noninvariance and obey spacetime uncertainty relation. We finally argue that, 
at high energy-momentum limit, the `electric' noncommutative gauge theory 
describes correctly dynamics of delocalized multiple F-strings.
\vspace*{1.1cm}

\baselineskip=18pt
\newpage

\section{Introduction}
\setcounter{equation}{0}
Following the recent understanding concerning the equivalence between
non-commutative and commutative gauge theories \cite{SW}, an 
immediate question to ask is how the theory behaves under the S-duality
interchanging the strong and weak coupling regimes. Indeed, this question has 
been addressed in several recent works \cite{ours,sst,harvard,prince,
sjrey} (See also related works \cite{rabinovici, gomis, omtheory, russo,
alvarezgaume, kawano}).

One expects an answer as simple as follows. The four-dimensional 
non-commutative theory describes the low-energy worldvolume dynamics 
of a D3-brane in the presence of a background of NS-NS two-form potential
(Kalb-Ramond potential) $B_2$ but none others. Under the S-duality of the 
Type IIB string theory, the D3-brane is self-dual, while the NS-NS two-form 
potential $B_2$ is swapped with the R-R two-form potential $C_2$. 
Thus, after the S-duality, the dual theory appears to be the one 
describing the low-energy worldvolume dynamics of a D3-brane in the
presence of a
R-R two-form potential $C_2$, but none others. Since there is no NS-NS
two-form potential present, noncommutative deformation via Seiberg-Witten 
map is not possible and the resulting theory ought to be a commutative theory.
However, this is apparently not the answer one obtains 
\cite{sst,harvard,sjrey}. 
Starting with a noncommutative U(1) gauge theory with coupling constant 
$g$ and noncommutativity tensor $\theta$ and taking the standard duality
transformation, one finds that the dual theory still remains a noncommutative 
U(1) gauge theory, but with coupling and noncommutativity parameters 
\be
g_{\rm D} = {1 \over g} \qquad {\rm and} 
\qquad \theta_{\rm D}  = - g^2 \widetilde{\theta},
\label{dualparm}
\ee
where $\widetilde{\theta}^{\mu \nu} = {1 \over 2} {\epsilon^{\mu \nu}}_{\alpha
\beta} \theta^{\alpha \beta}$. Alternatively, as is done in \cite{sst, 
harvard}, one may utilize the gauge invariance of $(F + B_2)$ to 
dial out the space-space noncommutativity and treat the theory as the 
standard gauge theory in a constant magnetic field. 
The S-duality would then turn this into a dual gauge theory, but now in a 
constant electric field, which is gauge equivalent to a theory with space-time
noncommutativity. Actually, in the dual theory, it turns out to be 
impossible to take a field theory limit that the dual theory would be 
best described by a noncritical open 
string theory, whose tension is of the order of the noncommutativity scale.

The aim of this paper is to understand the S-duality via the following routes:
\be
\begin{array}{ccc}
{\rm D3-brane} & \longleftrightarrow & \widetilde{\rm D3}-{\rm brane} \\
\uparrow & & \uparrow \\
\vert & & \vert \\
\vert & & \vert \\
\downarrow & & \downarrow \\
{\rm NCYM} & \longleftrightarrow & \widetilde{\rm NCYM} \\
\end{array}
\nonumber
\ee
and, if possible, to reconcile
these seemingly different results concerning the S-duality of the 
noncommutative gauge theory from various perspectives. 

\section{Naive S-Duality}
In this section, we shall be studying S-duality of noncommutative U(1)
gauge theory \footnote{This section is based on the result of \cite{ours}. 
Later, similar but independent derivation of the first half of this section 
has appeared in \cite{prince}.}.
The latter is defined by the following action:
\be
S &=& {1 \over 4g^2}  \int \! d^4 x \, \tr \Big( {\cal F} \star {\cal F} \Big) ,
\label{action}
\ee
where the noncommutative field strength is defined by  
\be
{\cal F} = \F + i \{ A_\mu, A_\nu \}_\star,
\qquad \F_{\mu \nu} = (\partial_\mu A_\nu - \partial_\nu A_\mu)
\nonumber
\ee
and $\tr$ refers to ``matrix notation'' for spacetime index contractions, 
for instance, 
$\tr\left( {\cal F} {\cal F} \right)$ means $ {{\cal F}_{\mu}}^{\nu} 
{{\cal F}_{\nu}}^{\mu}$. 
Seiberg and Witten have found explicitly the map between ${\cal F}_{\mu \nu}$ 
and $\F_{\mu \nu}$ :
\be 
{\cal F} = \F -  \left[ \F \theta \F  - (A \theta \partial) \F \right] + \cdots .
\label{swmap}
\ee
The map Eq.(\ref{swmap}) allows to expand the action 
Eq.(\ref{action}) in powers of dimensionless combination $\theta \F$: 
\be
S = \frac{1}{4g^2} \int d^4 x \, 
\left[-\tr \left( \F \F \right) + 2 \tr \left( \theta 
\F \F \F \right) - \frac{1}{2} \tr \left( \theta \F \right) \tr \left( 
\F \F \right) + \cdots \right].
\label{origexp}
\ee
Adopting the standard rule of duality transformation, we can promote the 
field strength $\F$ (not ${\cal F}$) into an unconstrained field by 
including a term $\widetilde{\G} \F$ where 
$\widetilde{\G}_{\mu\nu} = \frac{1}{2}
\epsilon_{\mu\nu}^{\;\;\;\;\rho\sigma} \G_{\rho\sigma}$ and $ \G_{\mu\nu}
= \partial_{\mu} B_{\nu} - \partial_{\nu} B_{\mu}$. 
Varying with respect to $\bf B$ imposes the constraint $d {\bf F} = 0$
which is solved by $ {\bf F} = d {\bf A}$ 
and we get back the original theory. On
the other hand we can now vary with respect to $\F$ which gives us
\be
 g^2\widetilde{\G} = {\bf F} - (\theta {\bf F} {\bf F} + 
{\bf F} \theta {\bf F} + {\bf F} {\bf F} \theta) 
+ \frac{1}{4} \tr \left( {\bf F} {\bf F} \right) \theta 
+ \frac{1}{2} \tr \left( \theta {\bf F} \right) {\bf F} + \cdots
\nonumber
\ee
which can be inverted, to the lowest order in $\theta$, as
\be
 {\bf F} = g^2\widetilde{\G} + g^4\theta \widetilde{\G} \widetilde{\G} 
+ g^4\widetilde{\G} \theta \widetilde{\G} 
+ g^4\widetilde{\G} \widetilde{\G} \theta 
- \frac{g^4}{4} \tr \left( \widetilde{\G}\widetilde{\G} \right) \theta 
- \frac{g^4}{2} \tr \left( \theta \widetilde{\G} \right) \widetilde{\G} 
+ \cdots.
\nonumber
\ee 
After the duality transformation, the action becomes
\be
S =  \frac{g^2}{4} \int \! d^4 x \, 
\left[ \tr \left( \widetilde{\G}\widetilde{\G} \right)
  + 2 g^2 \tr \left(\theta \widetilde{\G} \widetilde {\G} \widetilde{\G} \right)
 -\frac{g^2}{2} \tr \left(\theta \widetilde{\G} \right)
  \tr \left( \widetilde{\G} \widetilde{\G} \right) + \cdots \right].
\nonumber
\ee
This can be rewritten as 
\be
S = \frac{1}{4 g_{\rm D}^{2}}\int \! d^4 x \, 
\left[ - \tr \left( \G \G \right) + 2 \tr \left( \theta_{D} \G \G \G \right)
 -\frac{1}{2}\tr \left( \theta_{D} \G \right)
  \tr \left( \G \G \right) + \cdots \right] ,
\label{dualexp}
\ee
where $g_{\rm D}$ and $\theta_{\rm D}$ are given as in Eq.(\ref{dualparm}). 
It is evident that the above action can be reorganized into a self-dual
form
\be
S = \frac{1}{4 g_{\rm D}^2} \int \! d^4 x \, 
\tr \Big( {\cal G} \widetilde{\star} {\cal G} \Big)
\label{dualaction}
\ee
when expanded in powers of $\theta_{\rm D} {\bf G}$ and expressed
noncommutativity via $\theta_{\rm D}$, where
\be
{\cal G}_{\mu \nu} = \G_{\mu \nu} + i \{ B_\mu, B_\nu \}_{\widetilde{\star}}, 
\qquad
\G_{\mu \nu} 
= \partial_\mu {\bf B}_\nu - \partial_\nu {\bf B}_\mu.
\nonumber
\ee
Thus, we find that the dual action Eq.(\ref{dualaction}) is again 
noncommutative U(1) gauge theory, but with dual coupling parameters
Eq.(\ref{dualparm}). Note here that if we started with a theory with 
space-space non-commutativity, the dual theory will have space-time 
non-commutativity.

The foregoing analysis may be extended to the situation where R-R zero-form 
$C$ and two-form $C_2$ background is turned on.
In this case, the full action takes the form:
\be
S_{\rm total} = S + 
\int \! d^4 x \, \left[ \frac{1}{2}C\tr\left(\F \widetilde{\F} \right)
 + \tr \left( {\bf F} \widetilde{C}_2 \right) + \cdots \right].
\nonumber
\ee
Here, one may wonder what form of the coupling one ought to use. One
could for instance imagine that the correct coupling should be given
by a standard form but in terms of the noncommutative field strength, 
for example, $C\tr\left({\cal F} \star {\widetilde{\cal F}}\right)$, etc. 
We have checked, again to the lowest order in noncommutativity parameter, 
that modifying the coupling in this fashion does not change the final result
we will draw.

It is obvious that the effect of the R-R two-form potential is simply to 
shift $ {\bf G} \rightarrow {\bf G} + C_2$ 
everywhere and, for the R-R zero-form, 
a straightforward calculation shows that the action is self-dual (i.e. of
the same form as the original one) under the duality transformation {\sl 
provided} that we define the dual parameters and backgrounds as: 
\be
 g_{D}^{-2} &=& \frac{g^{-2}} {\left(g^{-4} + C^2 \right)}\nonumber\\
 C_{D} &=& -\frac{C }{\left(g^{-4} + C^2\right)}\\ 
 \theta_{D} &=& C_{D}\theta -\frac{1}{g_{D}^{2}}\tilde{\theta}.
\nonumber
\ee
These are the results anticipated from Type IIB S-duality:
\be
{\rm S}_{\rm IIB} : \qquad \quad \left(C + { i  \over g^2} \right) 
&\rightarrow& - \left( C + {i \over g^2} \right)^{-1}
\nonumber \\
\left( \theta + i \widetilde{\theta} \right) &\rightarrow& 
\left(C + {i \over g^2} \right)^{-1} \left(\theta + i \widetilde{\theta} 
\right).
\ee
Note that, had we started with an original theory having purely `electric'
or `magnetic' noncommutativity, viz. $\theta$ is a tensor of rank-1, in the
presence of the R-R scalar background, the dual theory would have 
noncommutativity tensor $\theta_{\rm D}$ of full rank.

\section{Closer Look - Noncritical Open String}
We have seen, in the previous section, that the naive S-dual of a
noncommutative gauge theory is again a noncommutative gauge theory, but 
with the noncommutativity parameter $\theta_{\rm D} = - g^2 \tilde{\theta}$. 
Thus, if the original theory were defined with `magnetic' noncommutativity, 
the dual theory would have `electric' noncommutativity and vice versa. 
We also note that, in performing the S-duality map, as we have expanded 
the original and the dual actions as in Eq.(\ref{origexp}) and 
Eq.(\ref{dualexp}), respectively, the result would be valid only for
$\theta {\bf F} \ll 1$ {\sl and} $\theta_{\rm D} {\bf G} \ll 1$. 
The latter requires that $g \ll 1$ and, in this case, the physics
in the original and the dual theories wouldn't be different much from the
standard commutative gauge theories \footnote{Recall that, for pure U(1)
gauge theories, the S-duality map is exact for {\sl all} values of $g$.}.

What can one say for the $g \gg 1$ case, viz. the original theory is strongly
coupled? In order to understand this limit, 
for definiteness, we will take the noncommutativity purely `magnetic':
$\theta = \theta_{23}$. The spectrum in this theory includes, in addition
to the U(1) gauge boson, the magnetic monopole and the dyon (See \cite{gross}
and references therein). They may be viewed as  noncommutative deformations of 
the photon, the magnetic monopoles and the dyons in the 
standard U(1) gauge theory. In noncommutative gauge theory, the U(1) gauge 
boson can be visualized as an induced electric dipole. 

\begin{figure}[t]
   \vspace{1cm}
   \epsfysize=6cm
   \epsfxsize=6cm
   \centerline{\epsffile{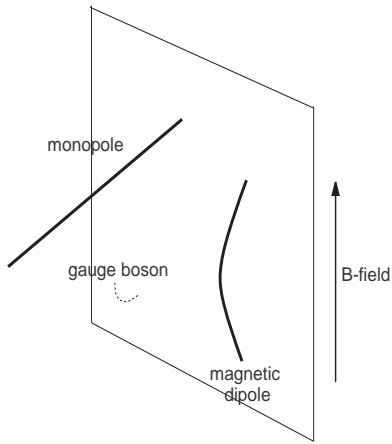}}
   \vspace*{.5cm}
\caption{\label{fig1} Cartoon view of gauge boson (open F-string dipole), 
Dirac magnetic monopole and magnetic dipole (open D-string dipole).}
\end{figure}

Alternatively, 
one may analyze the spectrum as fundamental (F-) or D-strings, respectively, 
ending on the D3-brane on which background magnetic field is turned on. The 
latter should be describable in terms of the Dirac-Born-Infeld Lagrangian 
($T_{\rm F} = 1/2 \pi \alpha'$):
\be
L_{\rm DBI} = - {T_{\rm F}^2 \over \lambda_{\rm st}}
\sqrt{-g_{00} g_{11}} \sqrt{g_{22} g_{33} + {\bf F}_{23}^2/T_{\rm F}^2 }
\nonumber
\ee
where, in the Seiberg-Witten decoupling limit, the bulk coupling parameter are 
related to the gauge theory parameters in Eq.(\ref{action}) as
\be
\lambda_{\rm st} = {g^2 \over T_{\rm F} \theta}, \qquad
-g_{00} = g_{11} = 1, \qquad
g_{22} = g_{33} = \left( {1 \over T_{\rm F} \theta} \right)^2, 
\qquad {\bf F}_{23} = \left( {1 \over \theta} \right) \epsilon_{23}.
\nonumber
\ee
The magnetic monopoles are not part of the physical spectrum as, being
represented by semi-infinite D-string ending on the D3-brane, they are
counterpart of the Dirac magnetic monopole with an infinite self-energy. 
Among the physical excitations, however, are {\sl magnetic dipoles} 
(as well as their dyonic counterparts) composed of monopole-antimonopole pair. 
See figure 1. 
Consider an open D-string (lying entirely on the D3-brane) of length $\Delta
{\bf x}$. It represents a magnetic dipole carrying a 
dipole moment ${\bf m} = \Delta
{\bf x}$ (measured in string unit) and total mass 
\be
M_{\rm dipole} = \left({ T_{\rm F} \over \lambda_{\rm st}}\right) \vert
\Delta {\bf x} \vert -  {\bf m} \cdot {\bf H}.
\label{energy}
\ee
where the negative sign in the second term refers to relative
{\sl opposite} orientation between the dipole and the background 
magnetic field. 
The last term represents the interaction energy of the magnetic dipole
with the background magnetic field: 
\be
{\bf H} = - {\partial L_{\rm DBI} \over \partial {\bf F}_{23}}
= {\bf H}_{\rm c} 
\left[ 1 + \left({1 \over T_{\rm F} \theta} \right)^2 \right]^{-1/2},
\label{magfield}
\ee
where
${\bf H}_{\rm c}$ denotes the critical magnetic field strength
\be
{\bf H}_{\rm c} = {T_{\rm F} \over \lambda_{\rm st}}.
\nonumber 
\ee
In the field theory limit $T_{\rm F} \rightarrow \infty$, the magnetic
dipole remain as low-energy excitaitons -- they are {\sl noncritical}
open D-strings with an effective tension
\be
T_{\rm eff} = { M_{\rm dipole} \over \vert \Delta {\bf x} \vert}
\quad \longrightarrow \quad {1 \over 2 g^2 \theta}.
\label{tension}
\ee

In Eq.(\ref{energy}), the negative sign in the second term refers to 
relative {\sl opposite} orientation between the dipole and the background 
magnetic field. It implies that the noncritical open D-string representing 
the magnetic dipole ought to be {\sl chiral}: open D-string with opposite
orientation, which represents magnetic {\sl anti}-dipole, is separated by 
an infinite mass gap from the magnetic dipoles. Thus, from the open D-string 
point of view, the field theory limit $T_{\rm F} \rightarrow \infty$ amounts 
to taking {\sl non-relativistic} limit and allows to expand Eq.(\ref{magfield})
in power-series of $(1/T_{\rm F} \theta)^2$. This also account for physical
origin of the numerical factor 1/2 in Eq.(\ref{tension}) \footnote{A related
observation was made recently by Klebanov and Maldacena in the context of
(1+1)-dimensional noncommutative open string theory \cite{KM}.}.

Taking the strong coupling limit, $g \gg 1$, unlike the magnetic monopoles,
the magnetic dipoles are nearly
tensionless, weakly interacting degrees of freedom, while the U(1) gauge
bosons are tensionless, strongly interacting degrees of freedom. Performing
S-duality Eq.(\ref{dualparm}) to the dual gauge theory, the two are 
interchanged with each other:
the {\sl dual electric dipoles} are nearly tensionless, weakly interacting 
degrees of freedom and the dual magnetic dipoles are tensionless but strongly
interacting degrees of freedom. They are made out of open F- and D-strings 
ending on the dual D3-brane worldvolume, on which a background dual 
{\sl electric} field ${\bf G}_{01}$ is turned on. The background 
dual electric field ought to be near critical, as is anticipated from 
S-duality and evidenced by the fact that, for fixed $\theta$, 
$\theta_{\rm D} = g^2 \theta$ becomes infinitely large. 
Indeed, combining the S-duality transformation of gauge theory parameters
Eq.(\ref{dualparm}) and of bulk coupling parameters
\be
\lambda_{\rm D, st} = {1 \over \lambda}_{\rm st} \qquad {\rm and} \qquad
g_{\rm D, \mu \nu} = {1 \over \lambda}_{\rm st} g_{\mu \nu},
\nonumber
\ee
we find that ${\bf F}_{23}$ is mapped to displacement field ${\bf D}
= -({\partial L_{\rm DBI} / \partial {\bf G}_{01}})$ and the dual
electric field ${\bf E}_{\rm D} := {\bf G}_{01}$ is given by
\be
{\bf E}_{\rm D} = {\bf E}_{\rm c} 
\left[ 1 + \left( {1 \over g_{\rm D}^2 T_{\rm F}
\theta_{\rm D}} \right)^2 \right]^{-1/2} \qquad
{\rm where}
\qquad
{\bf E}_{\rm c} = g^4_{\rm D} T_{\rm F}^2 \theta_{\rm D}.
\nonumber  
\ee
Likewise, magnetic dipoles are mapped into electric
dipoles made out of open F-strings, whose effective tension is given by
the S-dual of Eq.(\ref{tension}):
\be
\widetilde{T}_{\rm eff} = {1 \over 2 \theta_{\rm D}},
\label{dualtension}
\ee
where again the factor of 1/2 ought to signify {\sl chiral} nature
of the open F-string. An immediate question is: are these nearly tensionless, 
open F-strings identifiable within the dual gauge theory?
 
A set of gauge invariant operators in the dual gauge theory is
given by the following open Wilson lines \cite{wilinc}:
\be
\widetilde{W}_{\bf k} [C] = \int \! d^2 {\bf x} \, 
\exp_{\widetilde\star} 
\Big[ i \int_C \, \dot y(t) \cdot {\bf B} (x + y(t)) \Big]
\,\widetilde{\star} \, e^{i {\bf k} \cdot {\bf x} }.
\label{wilsonline}
\ee
Here, $t=[0,1]$ denotes the affine parameter along the open Wilson
line, $x^\mu$ refers to the spacetime position of the $\tau = 0$ point, 
${\bf x}$ to the projection of $x^\mu$ onto the two-dimensional noncommutative 
spacetime and ${\bf k}$ to the Fourier-momentum along the noncommutative
spacetime. All 
the multiplications are defined in terms of the `generalized Moyal product':
\be
A(x) \, \widetilde{\star} \, B(y) \,\, := \,\, 
\exp \left( {i \over 2} \theta^{\mu \nu}_{\rm D}
\partial_\mu^x \partial_\nu^y \right) A(x) B(y).
\nonumber
\ee 
It is then straightforward to check that the Wilson line Eq.(\ref{wilsonline})
is gauge invariant
{\sl provided} the following relation holds between the momentum ${\bf k}_\mu$
and the endpoint separation distance:
\be
{\bf k}_\mu = \left({ 1 \over \theta_{\rm D}} \right)_{\mu \nu} \Delta y^\nu
\qquad \,\,\, {\rm where} \qquad \,\,\,
\Delta y \equiv y(1) - y(0).
\label{uvir}
\ee

\begin{figure}[t]
   \vspace{1cm}
   \epsfysize=6cm
   \epsfxsize=6cm
   \centerline{\epsffile{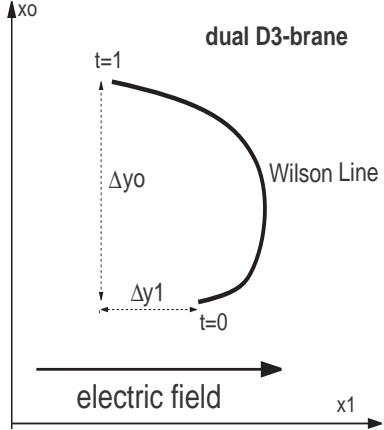}}
   \vspace*{.5cm}
\caption{\label{fig2} Characteristic shape of Wilson line.}
\end{figure}

What happens here is that, once the relation Eq.(\ref{uvir}), 
the extra phase factor $e^{ i {\bf k} \cdot {\bf x}}$ effectively 
parallel transports the gauge transformation parameter at $t = 1$ back to 
that at $t = 0$. This then ensures that the noncommutative Wilson line, 
despite being an open string, is gauge invariant. Being so, much as
the closed Wilson loops form a complete set of gauge invariant observables
in Yang-Mills theory, we can take the noncommutative Wilson lines 
Eq.(\ref{wilsonline}) as a complete set of gauge invariant observables of 
the dual noncommutative U(1) gauge theory \footnote{For noncommutative 
U(N) gauge theory, via Morita equivalence, we expect that the noncommutative 
Wilson lines Eq.(\ref{wilsonline}) still form a complete set of gauge 
invariant observables.}. 

For small total energy or momentum, 
$\vert {\bf k} \vert \ll 1/\sqrt{\theta_{\rm D}}$, separation between the 
Wilson line endpoints is shorter than the non-commutativity scale,
$\vert \Delta y \vert \ll \sqrt{\theta_{\rm D}}$. 
Hence, the Wilson line reduces
effectively to a (Fourier-transform) of the standard closed Wilson loop.
For energies smaller than the non-commutativity scale 
$1/{\sqrt{\theta_{\rm D}}}$, we would expect the dual theory behaves as 
in the standard gauge theory. This agrees with the conclusions of \cite{sst, harvard, 
sjrey}. On the other hand, if $\vert {\bf k} \vert \gg 
1/\sqrt{\theta_{\rm D}}$, then the separation of the Wilson line endpoints 
would be larger than the noncommutativity scale, 
$\vert \Delta y \vert \gg \sqrt{\theta_{\rm D}}$. These excitations are
string-like. 

Taking the zeroth component of Eq.(\ref{uvir}), one finds
\be
E_{\rm YM} \,\, := \,\, \widetilde{T}_{\rm eff} \, \Delta y^1  , 
\label{YMenergy}
\ee
where $\widetilde{T}_{\rm eff}$ coincides precisely with the effective tension
Eq.(\ref{dualtension}).
Thus, we are prompted to identify the Wilson line with a noncritical
open F-string, whose effective tension is given by $\widetilde{T}_{\rm eff}$.
Recall that, under the S-duality Eq.(\ref{dualparm}), 
$T_{\rm eff} = 1/2 g^2 \theta$ is mapped to 
$\widetilde{T}_{\rm eff} = 1/2 \theta_{\rm D}$. Hence, the
open Wilson lines in the dual gauge theory are the right candidates for
the S-dual of the magnetic dipoles in the original gauge theory. 

Characteristic size of the
open Wilson lines is ${\cal O}(\sqrt{\theta_{\rm D}})$ and become 
macroscopically large in the strong coupling limit $g \gg 1$. They represent
a complete set of excitations in the weakly coupled, dual gauge theory 
for $\vert {\bf k} \vert \ge 1/\sqrt{\theta_{\rm D}} \sim 0$. 
It clearly suggests that the noncommutative gauge theory captures more of the
description than we had the right to expect from the naive duality argument
in Section 2.
It also indicates that a suitable formulation of the dual gauge theory is 
in terms of macroscopic open strings. 

In extracting tension of the open Wilson line from Eq.(\ref{YMenergy}),
we were able to match it to Eq.(\ref{dualtension}) {\sl modulo} a numerical
factor of 2. Recall that the numerical factor of 1/2 in Eqs.(\ref{tension}, 
\ref{dualtension})
has originated from {\sl chiral} or, equivalently, {\sl non-relativistic}
nature of the open D- and F-strings. What then would cause the open Wilson
lines of the noncommutative Yang-Mills theory {\sl chiral} and eventually 
account for cancellation or disappearance of the factor of 2 in Eq.
(\ref{YMenergy})?

We believe an answer to this question comes from the fact that `electric'
noncommutative Yang-Mills theory is parity-violating. For fixed 
$\theta_{\rm D}$, this is easily seen from non-invariance of the electric
noncommutativity $[x^0, x^1] = \theta_{\rm D}$ under ${\bf x} \rightarrow
- {\bf x}$ \footnote{Note, however, that `magnetic' noncommutative Yang-Mills 
theory is parity-conserving: magnetic noncommutativity $[x^2, x^3] = \theta$
is invariant under ${\bf x} \rightarrow - {\bf x}$.}. 
This then implies that, along the $(x^0-x^1)$-directions along which 
${\bf k}_\mu$ and $\Delta y^\mu$ in Eq.(\ref{uvir}) point,
the open Wilson lines ought to be {\sl chiral}, 
stretching the two endpoints such that $\Delta y^1$ positive. 
This chirality also implies that, from Eq.(\ref{YMenergy}), only positive
energy Wilson lines are physical excitations but not negative energy ones. 
The net result is essentially the same as that of the infinite mass gap     
opening up in the non-relativistic limit. 

There exists one more piece of evidence that the Wilson lines are indeed
identifiable with a sort of macroscopic string. Utilizing 
$E \sim \Delta E \ge \hbar 
/\Delta y^0$, we observe that the Wilson line exhibits 
a version of the {\sl spacetime uncertainty relation}:
\be
\Delta y^0 \Delta y^1 \,\, \ge \,\, {1 \over 2} \hbar \theta_{\rm D}.
\nonumber
\ee
As emphasized by Yoneya \cite{yoneya}, 
the spacetime uncertainty relation is a distinguishing 
feature of any string theory with worldsheet conformal invariance.

\section{Yet Another Look -- Strong Noncommutativity Limit}
We would like to discuss yet another piece of physics associated with the 
S-duality, Eq.(\ref{dualparm}). In the previous section, we have seen that,
due to the dual electric field background, the open F-string is oriented 
predominantly along the $x^0 - x^1$ directions. See Eq.(\ref{uvir}). There,
we have also argued that the open string is macroscopically stretched. 
According to Eq.(\ref{wilsonline}), the open string is made out of the
dual
gauge field ${\bf B}_\mu$ as a sort of coherent state configuration. Thus,
it ought to be possible to visualize the open string out of the dual gauge
theory in the semiclassical limit. In this section, under suitable 
condition, we show that the dual gauge theory Eq.(\ref{dualaction}) 
describes {\sl worldsheet} dynamics of $N_{\rm F}$ coincident macroscopic
F-strings propagating in four dimensional spacetime. 

Let us begin with the following elementary observation. Strongly coupled
noncommutative U(1) gauge theory with a {\sl finite} noncommutativity
$\theta$, as is seen from Eq.(\ref{dualparm}), is dual to weakly coupled
noncommutative U(1) gauge theory with an {\sl infinite} noncommutativity
$\theta_{\rm D}$. In dimensionless measure, this implies that
\be
\theta_{\rm D} \{ {\bf B}_\mu, {\bf B}_\nu \} \,\, \sim \,\, 
\theta_{\rm D} \big( \partial {\bf B} \big) \,\, \gg \,\, 1
\label{limit}
\ee
and hence corresponds to high field-strength, high-energy limit 
\footnote{This is the limit considered originally by \cite{largenc}}.
Because of the infinitely large noncommutativity, dynamics of the dual 
`electric' U(1) gauge theory is considerably simplified.
At leading order in the noncommutativity Eq.(\ref{limit}), 
the dual gauge theory action Eq.(\ref{dualaction}) is reduced as:
\be
S \quad \rightarrow \quad {1 \over 4 g_{\rm D}^2} V_\perp
\int dx^0 d x^1 \Big<\,\, \left( {\cal G}_{\mu \nu} \right)^2_{\widetilde 
\star} \,\, \Big>,
\nonumber
\ee
where the Lorentz indices are contracted with gauge theory metric. 
We have also introduced notations 
\be
V_\perp \equiv \int dx^2 dx^3 \qquad
{\rm and} \qquad 
\Big< \,\, {\cal O} \,\, \Big> \equiv
{1 \over  V_\perp} \int dx^2 dx^3 {\cal O}. 
\ee

In the limit of infinitely many coincident noncommutative D3-branes, it is 
known that 
nonabelian generalization of the dual gauge theory Eq.(\ref{dualaction}) may 
be interpreted as a theory describing low-energy dynamics of (F1-D3) bound 
states \cite{ppm2, harvard, harmark}. It
is known that, in this case, the Yang-Mills gauge coupling is not arbitrary 
but is determined by the F-string and the D3-brane charges 
$N_{\rm F}, N_{\rm D3}$
\cite{harmark}:
\be
g_{\rm D}^2 = {V_\perp \over \theta_{\rm D}} \left({ N_{\rm D3} \over
N_{\rm F}} \right). 
\nonumber
\ee 
Thus, using this parameter relation and introducing new covariant operator 
variables
\be
X^\mu = \theta^{\mu \nu}_{\rm D} \left( i \partial_\nu + {\bf B}_\nu \right),
\nonumber
\ee
we can re-express the dual `electric' noncommutative Yang-Mills theory 
action as:
\be
S = - {N_{\rm F} T_{\rm eff} \over 2}
\int dx^0 dx^1 \left( -{1 \over 2 }
\left< \,\, 
\left( {1 \over \theta_{\rm D}} 
\{ X^\mu, X^\nu \}_{\widetilde \star} \right)^2 \,\, \right> + \cdots \right)
\label{finalaction}
\ee
where again the Lorentz indices are contracted with respect to the 
gauge theory metric and the ellipses denote ${\cal O}(1)$ sub-leading terms. 
We have thus found that the resulting action Eq.(\ref{finalaction}) has 
precisely the same form as string worldsheet action for $N_{\rm F}$ multiple
noncritical, F-strings of tension $T_{\rm eff}$, except that the action 
is expressed in the so-called Schild gauge \cite{schild}. The F-strings 
ought to be interpreted as {\sl open} strings, albeit infinitely stretched, 
as they propagate in a spacetime governed by the gauge theory metric. Note 
that the worldsheet direction is along $(x^0-x^1)$-directions and the strings
are delocalized along $(x^2-x^3)$-directions.

Actually, what we have gotten is not precisely the Schild-gauge action but 
a deformation quantization of it. Namely, plaquette element 
$\Sigma^{\mu \nu}$ of the string worldsheet is deformed into
\be
\Sigma^{\mu \nu} \equiv
\epsilon^{ab} \partial_a X^\mu \partial_b X^\nu
= \{ X^\mu, X^\nu \}_{\rm PB}
\qquad 
\rightarrow \qquad
\Sigma^{\mu \nu}_{\widetilde \star} \equiv 
{1 \over \theta_{\rm D}} \{X^\mu, X^\nu \}_{\widetilde \star}.
\ee
We trust the deformation is correctly normalized, as 
$\Sigma^{\mu \nu}_{\widetilde \star} = \Sigma^{\mu \nu} + {\cal O}(\theta_D)$
in small $\theta_{\rm D}$ limit.

What conclusion can one draw out of the above result? Had we considered
$N_3$ coincident D3-branes with $N_{\rm F}$ units of center-of-mass U(1)
electric flux turned on, we would have obtained the standard Nambu-Goto
or Schild action of $N_{\rm F}$ F-strings. Recalling that noncommutative U(1)
gauge theory is equivalent to $\rm U(\infty)$ gauge theory at high-energy
regime, we may interpret that the dual `electric' noncommutative
gauge theory indeed describes worldsheet dynamics of $N_{\rm F}$ coincident 
F-strings {\sl provided} the latter carry high energy-momentum and become 
{\sl open} strings (see Eq.(\ref{uvir})), and are delocalized along 
$(x^2-x^3)$-directions. The result seems consistent with 
what one finds from supergravity dual \cite{harvard, harmark}.

\vskip1cm
\centerline{\bf Acknowledgement}
We thank M.R. Douglas, J. Kluso\v{n}, U. Lindstr\"{o}m and G. Moore
for useful discussions. SJR thanks warm hospitality of Marsaryk University, 
New High-Energy Theory Center at Rutgers University, and Theory Division
at CERN during completion of the work. 

\vskip0.5cm


\end{document}